\documentclass{vgtc}                          

\graphicspath{{figures/}{pictures/}{images/}{./}} 

\usepackage{times}                     

\usepackage{tabu}                      
\usepackage{booktabs}                  
\usepackage{lipsum}                    
\usepackage{mwe}                       

\usepackage{mathptmx}                  

\usepackage{colortbl}
\usepackage{xcolor}

\newcommand\tool{VRARE}
\newcommand\toolname{VR for Accessibility Requirements Engineering}
\onlineid{0}

\vgtccategory{Research}

\vgtcinsertpkg

\title{\tool: Using Virtual Reality to Understand Accessibility Requirements of Color Blindness and Weakness}

\author{Yi Wang\thanks{e-mail: xve@deakin.edu.au}\\ %
        \parbox{1.7in}{\scriptsize \centering School of Information Technology,  \\ Deakin University}%
\and Ben Cheng\thanks{e-mail: chengye@deakin.edu.au}\\ %
     \parbox{1.7in}{\scriptsize \centering School of Information Technology,  \\ Deakin University} %
\and Xiao Liu\thanks{e-mail: xiao.liu@deakin.edu.au}\\ %
     \parbox{1.7in}{\scriptsize \centering School of Information Technology,  \\ Deakin University}
\and Chetan Arora\thanks{e-mail: chetan.arora@monash.edu}\\ %
     \parbox{1.7in}{\scriptsize \centering Faculty of Information Technology,  \\ Monash University}
\and John Grundy\thanks{e-mail: john.grundy@monash.edu}\\ %
     \parbox{1.7in}{\scriptsize \centering Faculty of Information Technology,  \\ Monash University}
\and Thuong Hoang\thanks{e-mail: thuong.hoang@deakin.edu.au}\\ %
     \parbox{1.7in}{\scriptsize \centering School of Information Technology,  \\ Deakin University}
     }

\teaser{
  \centering

    \includegraphics[width=0.85\linewidth]{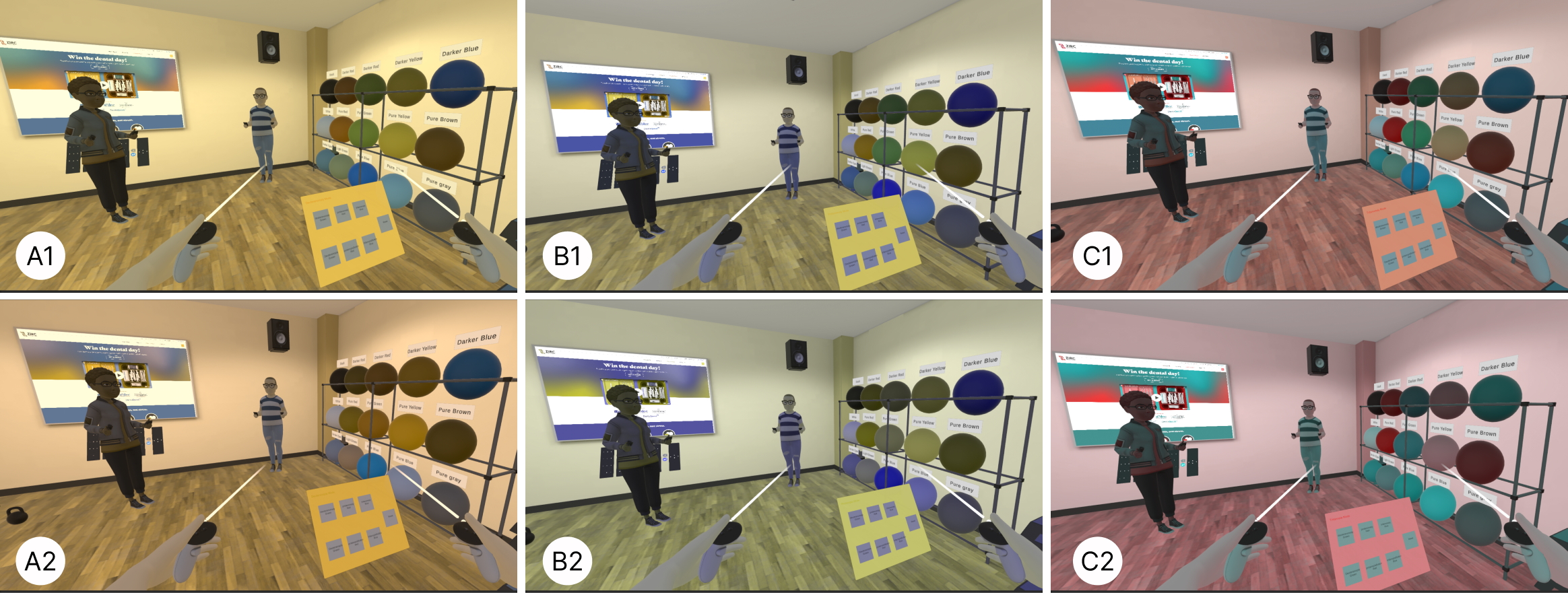}
  \caption{Teaser figure shows six types of color blindness: (A). A1-C1 represent Deuteranomaly, Protanomaly, and Tritanomaly, respectively. (B). A2-C2 represent Deuteranopia, Protanopia, and Tritanopia, respectively. }
  \label{fig:teaser}
}

\abstract{
   
In this paper, we developed a virtual reality (VR) system that can simulate color blindness and weakness. We built an immersive 3D web view interface where participants can discuss accessibility requirements for a fitness website projects within a virtual fitness environment. We conducted a pilot experiment involving 24 participants from six software teams, who used both VR and non-VR methods to understand color blindness and weakness requirements in a website project. Our findings indicate that using VR can provide several benefits for requirements activities, such as an improved user experience and reduced workload. 

} 


\begin{document}



\maketitle

\section{Introduction}


VR is a novel tool for capturing user requirements, including users with special needs \cite{wang2022vr4hcre_original, Wang2025}. VR provides unique features for Requirements Engineering (RE), including remote multiplayer collaboration, simulation of special user conditions, and various virtual environments (VEs). Immersive experiences may facilitate collaboration and even expressiveness among stakeholders. However, it is currently unclear which aspects of RE activities can be enhanced using VR, as previous work has only emphasized the potential value and novel ideas of using VR for RE, without exploring specific RE tasks. To fill this research gap, our study focuses primarily on using VR to simulate color vision deficiency (CVD), including color blindness and color weakness, helping participants understand accessibility requirements. We developed a novel VR system, \tool~ (\toolname), to investigate differences in workload and user experience between a VR-based approach (\tool) and a non-VR approach (face-to-face discussion) when discussing and understanding color blindness and CVD related requirements. Our results indicate that the VR method was associated with lower workload and a better user experience compared to non-VR methods.






\section{Related Work}

Recent studies have further reinforced VR's potential in this simulating visual impairment. Auslan Alexa – Vithanage et al. \cite{10108568} employed a VR Wizard-of-Oz prototyping approach to elicit design requirements from Deaf participants. This study demonstrates how VR’s immersive capabilities facilitate the capture of nuanced user feedback. BlueVR – You et al. \cite{10.1145/3611031} designed a VR serious game to simulate the experiences of individuals with color vision deficiency (CVD). By engaging non-CVD users in tasks that mimic the challenges faced by people with CVD, BlueVR enhances understanding and empathy, effectively addressing the limitations of traditional simulation methods. 

\section{\tool~ Design}



We implemented simulations of color blindness and color vision deficiency using Unity’s Post-Processing plug-in, applying Color Grading–based RGB channel blending adjustments to support the identification of visual accessibility requirements in VR (see Figure \ref{fig:teaser}). Complete CVD, including Deuteranopia, Protanopia, and Tritanopia, were simulated using established RGB channel transformation models, whereas color vision deficiencies with partial impairment (i.e., Deuteranomaly, Protanomaly, and Tritanomaly) were simulated through fine-grained adjustments of channel weights. 

\tool~was implemented in Unity3D (version 2020.3.20f1) and deployed on Meta Quest 3 with dual controllers. The remote collaboration functionality supports real-time voice communication and synchronized avatar behaviors to facilitate cross-location discussions. To enable color vision simulation on web-based content, we integrated a 3D Web View and made minor modifications to its rendering pipeline, allowing the color blindness and color blindness simulators to be consistently applied across both 3D scenes and embedded web interfaces.

\section{Pilot Experiment}

We conducted a pilot experiment to evaluate the impact of two methods on discussing and understanding accessibility requirements: VR method (\tool) and non-VR method.


\subsection{Participants}

We recruited 24 participants (18 males and 6 females) from the local campus, with ages ranging from 18 to 36 years old (\textit{Mean} = 24.8, \textit{SD} = 5.34). All participants had a background in software engineering and were actively engaged in undergraduate software projects. Among them, 18 (75\%) participants had prior experience with VR headsets, and 6 (25\%) participants had never used a VR headset. 14 (58\%) participants had often considered accessibility in their software projects; 7 (29\%) participants had an understanding of accessibility but did not consider it in their software projects and assignments; 3 (13\%) participants were unaware of accessibility. All participants were not color blind.

\subsection{Experiment Procedure}

We used a ``within-subjects'' design and designed discussion tasks: \textbf{1. Introduction and Training (5 minutes).} Participants were welcomed and briefed on the study’s purpose, data privacy, and duration, and were shown open-source accessible fitness website examples to establish a shared baseline understanding. Participants were randomly assigned to start with either the VR or the non-VR method. Six teams of four participants were formed. Teams in the VR method were each equipped with four Meta Quest 3 headsets and had access to the color vision deficiency simulator within the VR-based remote collaboration environment, whereas teams in the non-VR method were provided with laptops pre-installed with accessibility simulation tools. \textbf{2. Discussion on a Fitness Website (15 minutes). }Participants explored a researcher-provided fitness website and engaged in exploratory discussions. In the VR method, discussions were conducted entirely within the virtual environment using \tool~, whereas in the non-VR method, participants explored the same website on laptops and discussed the content face-to-face. No color vision or accessibility simulation tools were used in either method. This stage focused on fostering exploration and developing a shared understanding of how the website supported users with visual impairments, particularly users with color vision deficiencies. \textbf{3. Fitness Website Evaluation Task (15-20 minutes). }Teams conducted a systematic accessibility evaluation of the fitness website using color vision simulation tools, with a focus on identifying accessibility issues relevant to users with color vision deficiencies. In the VR method, teams evaluated the website within the virtual environment using \tool~, with color blindness and color vision deficiency simulations enabled. In the non-VR method, the same website was evaluated on laptops using a desktop-based color vision deficiency simulator. Participants collaboratively discussed the identified accessibility issues and potential improvements. After a short break, teams switched methods and repeated the task to enable a within-subjects comparison. \textbf{4. Debriefing.} Participants completed the NASA Task Load Index (NASA LTX) and User Experience Questionnaire (UEQ).

\section{Results and Discussion}\label{results}








\begin{figure}[t]
 \centering 
 \includegraphics[width=0.9\linewidth]{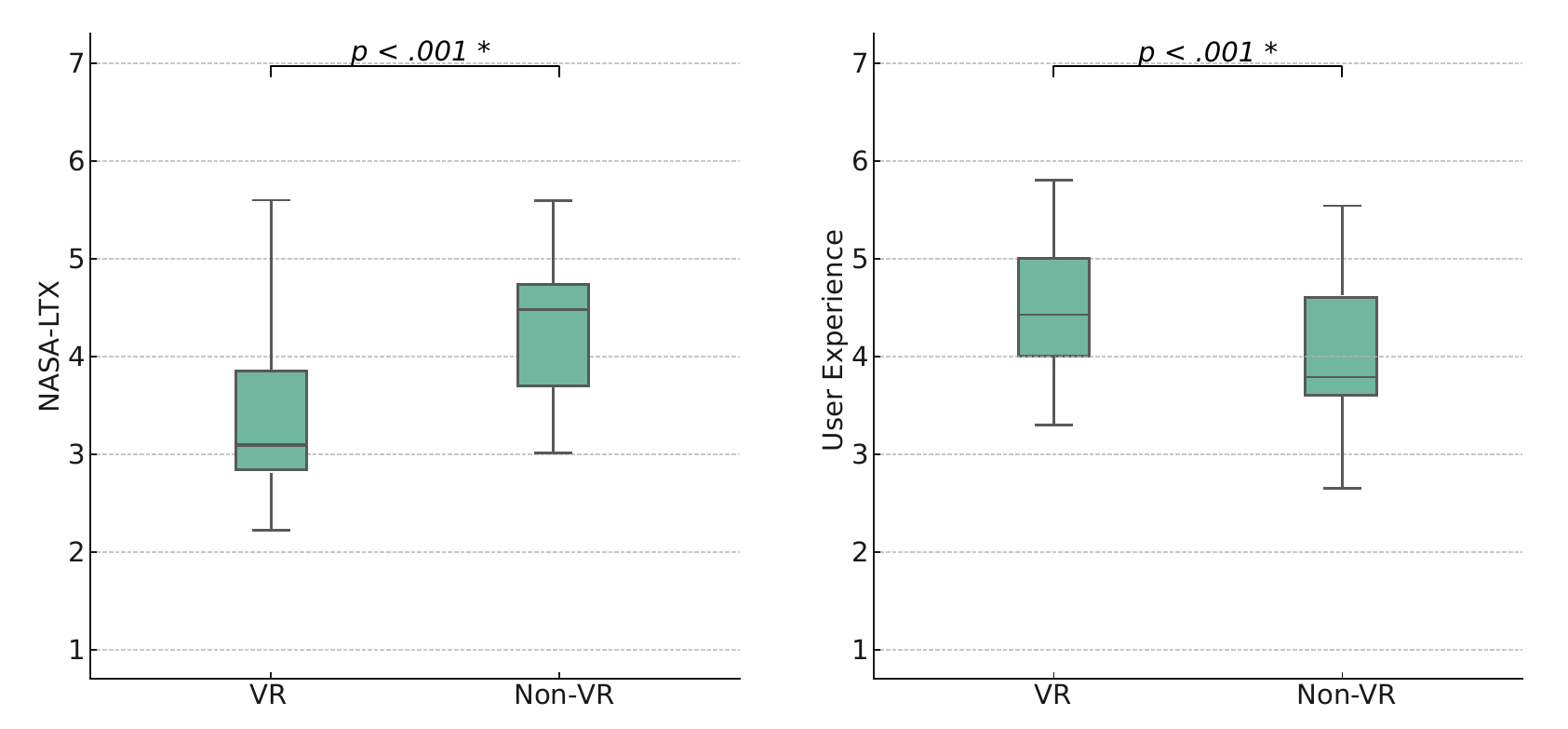}
 \caption{The results of the Workload and User Experience. }
 \label{fig:2}

\end{figure}

The Wilcoxon signed-rank test was used for data analysis. Figure \ref{fig:2} shows the results of the workload and user experience. For workload, there was a significant difference on workload between the VR method and the non-VR method (\textit{r} = 0.39, \textit{p} $<$ .001). Descriptive statistics indicated a lower score on workload when using the VR method. Thus, the workload associated with the VR-based approach was lower than that of the non-VR approach when discussing and understanding accessibility requirements. Employing VR to understand visual impairment–related characteristics may represent an effective approach in future work.

For user experience, there was a significant difference on user experience between VR method and non-VR method (\textit{r} = 0.46, \textit{p} $ < $ .001). Descriptive statistics indicated a higher score on user experience when using the VR method. Overall, the VR method provided a better user experience, which may be attributed to its higher level of immersion and more engaging interaction experience.

\section{Conclusion and Future Work }

We explored the differences between using VR method and non-VR method in understanding and discussing accessibility requirements. We developed \tool, a VR system that can simulate color blindness and weakness within the virtual environment. \tool~ also supports remote multi-user collaboration. Participants discussed accessibility requirements for a fitness website through a 3D Web View within the virtual environment. We conducted polit experiment to evaluate the differences between using VR method and the non-VR method in terms of workload.

In the future, we plan to further explore the role of VR in inspiring and analyzing accessibility requirements, and to evaluate how accessibility requirements captured through VR differ from those identified using traditional approaches. We also plan to develop gamified approaches to support RE activities and to explore their potential benefits.





\bibliographystyle{abbrv-doi}

\bibliography{template}
\end{document}